\documentclass[letter]{aa}

\usepackage{amsmath}
\usepackage{graphicx,txfonts,natbib}
\usepackage{longtable}
\usepackage{lscape}

\newcommand{\teff}{\mbox{${T}_{\rm eff}$}}
\newcommand{\logg}{\mbox{${\log g}$}}

\begin{document}
\title{Periodic mass-loss episodes due to an oscillation mode with variable
  amplitude 
in the hot supergiant HD\,50064\thanks{Based on high-resolution
spectroscopy assembled with the CORALIE spectrograph attached to the 1.2m Euler
telescope at La Silla, Chile and on CoRoT space-based photometry. The CoRoT
space mission was developed and is operated by the French space agency CNES,
with the participation of ESA's RSSD and Science Programmes, Austria, Belgium,
Brazil, Germany, and Spain.}}

\author{C.~Aerts\inst{1,2}
\and K.~Lefever\inst{1,3}
\and A.~Baglin\inst{4}
\and P.~Degroote\inst{1} 
\and R.~Oreiro\inst{1}
\and M.~Vu\v{c}kovi\'c\inst{1}
\and K.~Smolders\inst{1}
\and B.~Acke\inst{1,}\thanks{Postdoctoral Fellow of the Fund for Scientific
    Research of Flanders (FWO), Belgium}
\and T.~Verhoelst\inst{1,}$^{\star\star}$
\and M~.Desmet\inst{1}
\and M.~Godart\inst{5}
\and A.~Noels\inst{5}
\and M.-A.~Dupret\inst{5}
\and M.~Auvergne\inst{3}
\and F.~Baudin\inst{6}
\and C.~Catala\inst{3}
\and E.~Michel\inst{3}
\and R.~Samadi\inst{3}
}

\institute{Instituut voor Sterrenkunde, K.U.Leuven, Celestijnenlaan 200D, B-3001
Leuven, Belgium 
\and 
Department of Astrophysics, IMAPP, University of Nijmegen, PO Box 9010,
6500 GL Nijmegen, The Netherlands
\and
Belgisch Instituut voor Ruimte Aeronomie (BIRA), Ringlaan 3, B-1180 Brussels,
Belgium
\and
LESIA, CNRS UMR8109, 
Universit\'e Pierre et Marie Curie, Universit\'e Denis
Diderot, Observatoire de Paris, 92195 Meudon cedex, France
\and
Institut d'Astrophysique et de G\'eophysique, Universit\'e de Li\`ege, All\'ee
du 6 Ao\^ut 17, B-4000 Li\`ege, Belgium
\and
Institut d'Astrophysique Spatiale, CNRS/Universit\'e Paris XI UMR 8617, F-091405
Orsay, France
}

\date{Received ; accepted}

\authorrunning{Aerts et al.}
\titlerunning{Periodic mass loss episodes due to an oscillation
in the supergiant HD\,50064}

\offprints{conny.aerts@ster.kuleuven.be}

\abstract{}{We aim to interpret the photometric and spectroscopic
variability  of the 
luminous blue variable supergiant HD\,50064 ($V=8.21$).}  {CoRoT space
photometry and follow-up high-resolution spectroscopy with a time base of
137\,d and 169\,d, respectively, was gathered, analysed, and interpreted using
standard time series analysis and light curve modelling methods, as well as
spectral line diagnostics.}  {The space photometry reveals one period of 37\,d,
which undergoes a sudden amplitude change with a factor 1.6. The pulsation
period is confirmed in the spectroscopy, which additionally reveals metal line
radial velocity values differing by $\sim 30\,$km\,s$^{-1}$ depending on the
spectral line and on the epoch. We estimate \teff$\sim$13\,500\,K,
\logg$\sim$1.5 from the equivalent width of Si lines.  The Balmer lines reveal
that the star undergoes episodes of changing mass loss on a time scale similar
to the changes in the photometric and spectroscopic variability, with an average
value of $\log\dot{\rm M}\simeq-5$ (in M$_\odot$\,yr$^{-1}$).  We 
{tentatively
interpret the
37\,d period as the result of a strange
mode oscillation.}}{}


\maketitle

%

\section{Introduction}

One of the goals of the asteroseismology programme (Michel et al.\ 2006) of the
CoRoT satellite (Baglin et al.\ 2006) is to explore the Hertzsprung-Russell
diagram (HRD) through uninterrupted time series of white-light photometry of
unprecedented precision. In this context, numerous non-radial pulsators of
various kind have been observed and analysed, among which massive stars on the
main sequence (e.g., Degroote et al.\ 2009; Neiner et al.\ 2009).  With the 
goals of 
mapping the uppermost part of the HRD and understanding the role of oscillations
in the mass loss of evolved massive stars, a hot supergiant was observed in the
seismology programme of the satellite.
\begin{figure*}[ht!]
\begin{center}
\rotatebox{270}{\resizebox{5.cm}{!}{\includegraphics{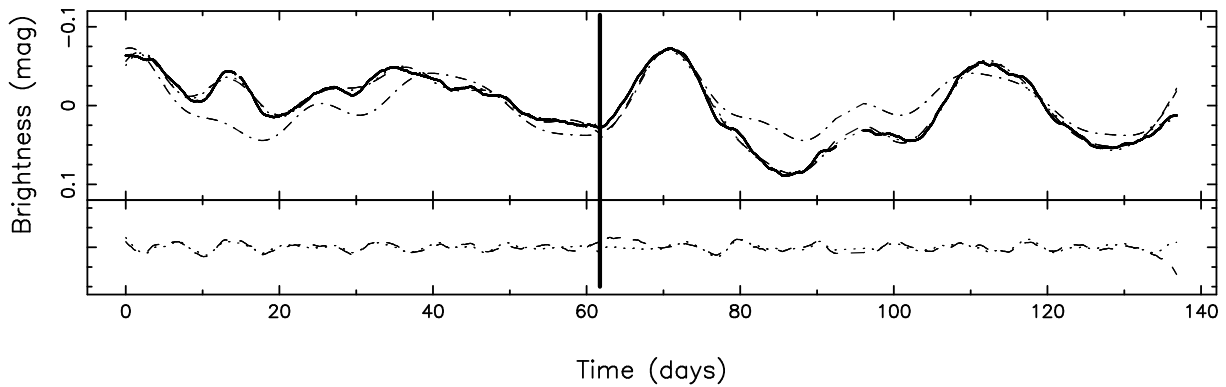}}}
\end{center}
\caption{The CoRoT light curve of HD\,50064 corrected for a linear downward
trend (full line). For an explanation of the fits in different linestyles, see
text. The $y$-axis of the lower panel is reduced by a factor 2 compared with the
one in the upper panel.}
\label{lc}
\end{figure*}
\begin{figure*}
\begin{center}
\rotatebox{270}{\resizebox{3.65cm}{!}{\includegraphics{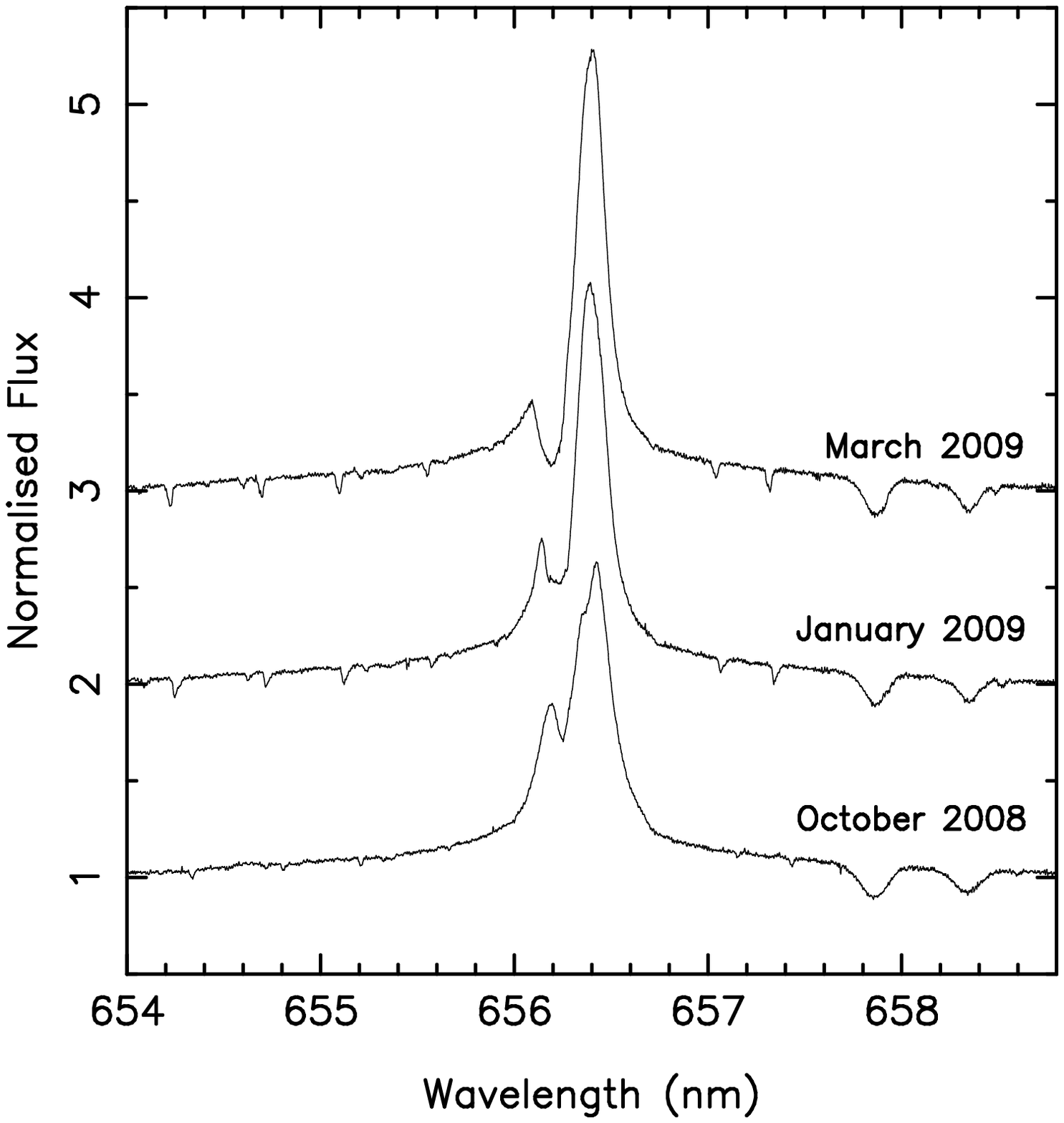}}}
\rotatebox{270}{\resizebox{3.65cm}{!}{\includegraphics{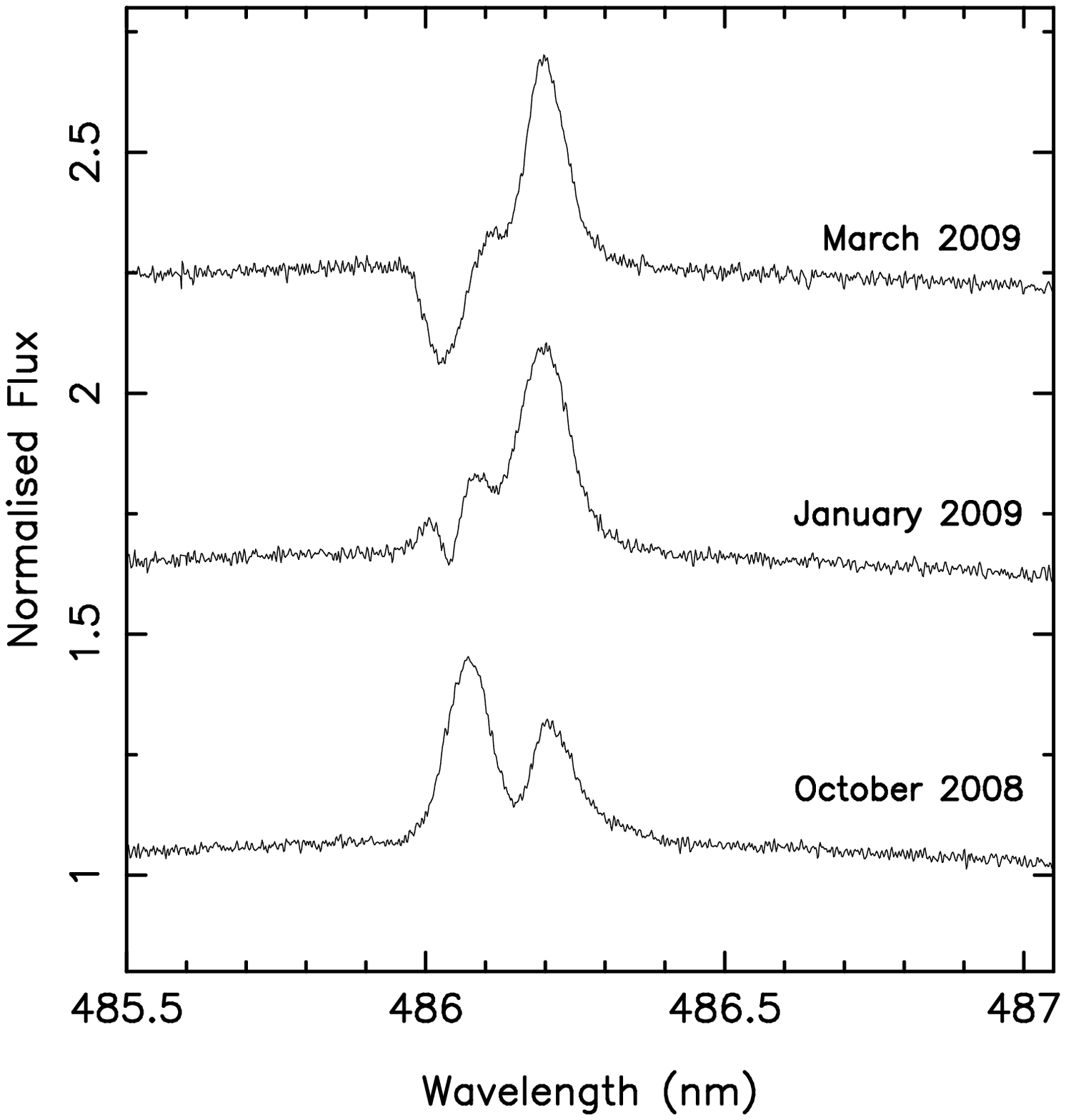}}}
\rotatebox{270}{\resizebox{3.65cm}{!}{\includegraphics{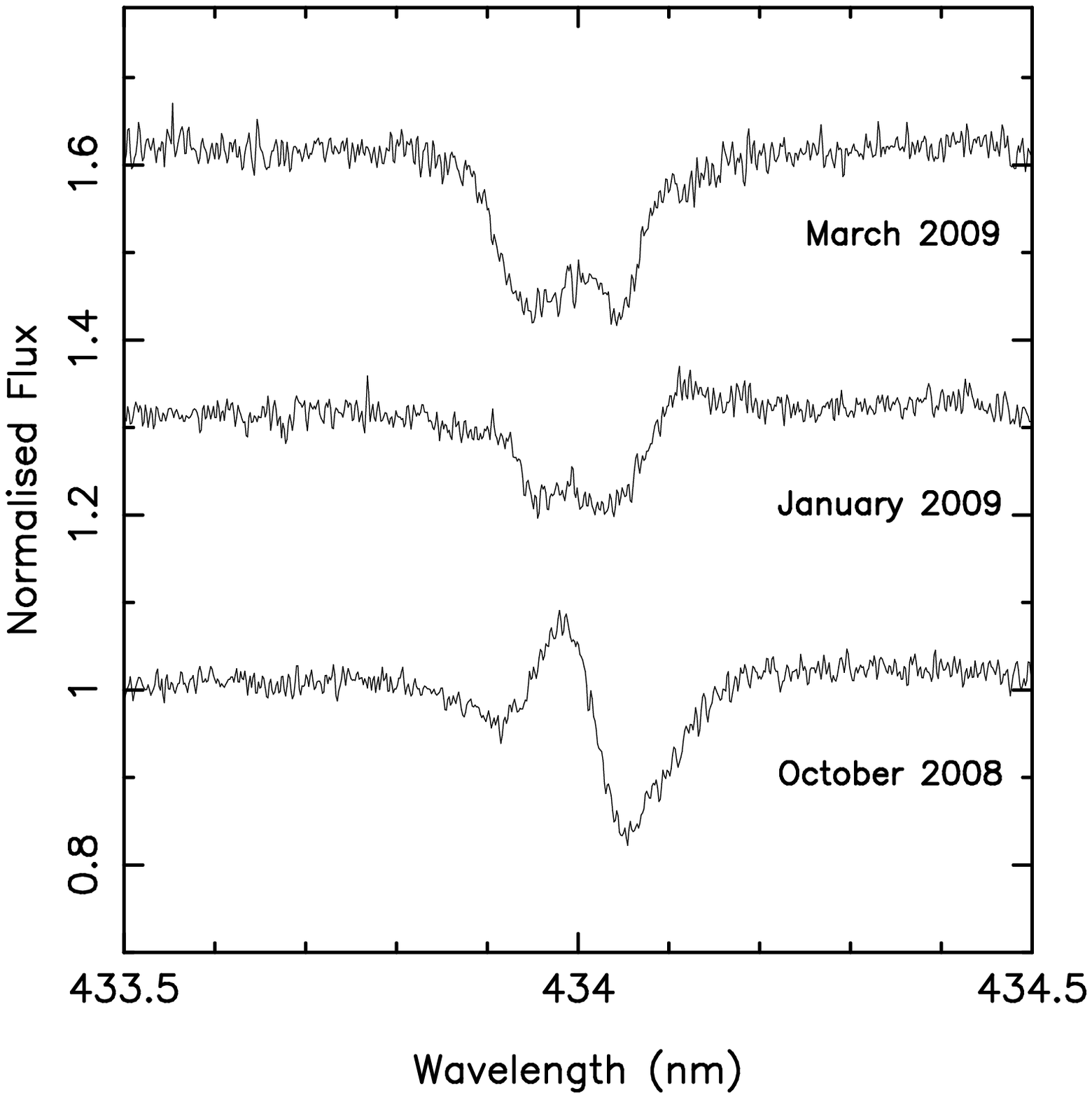}}}
\rotatebox{270}{\resizebox{3.65cm}{!}{\includegraphics{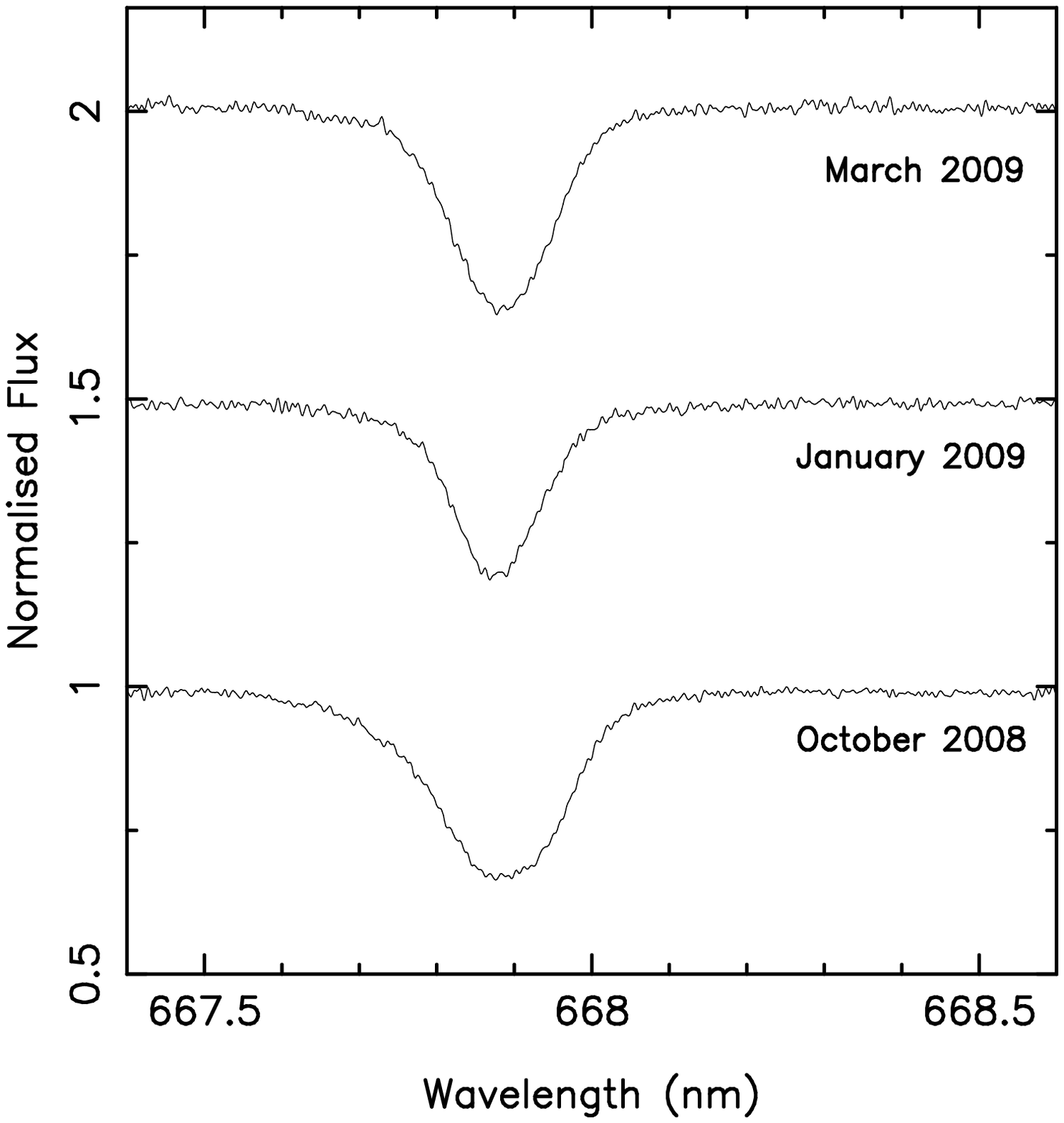}}}
\rotatebox{270}{\resizebox{3.65cm}{!}{\includegraphics{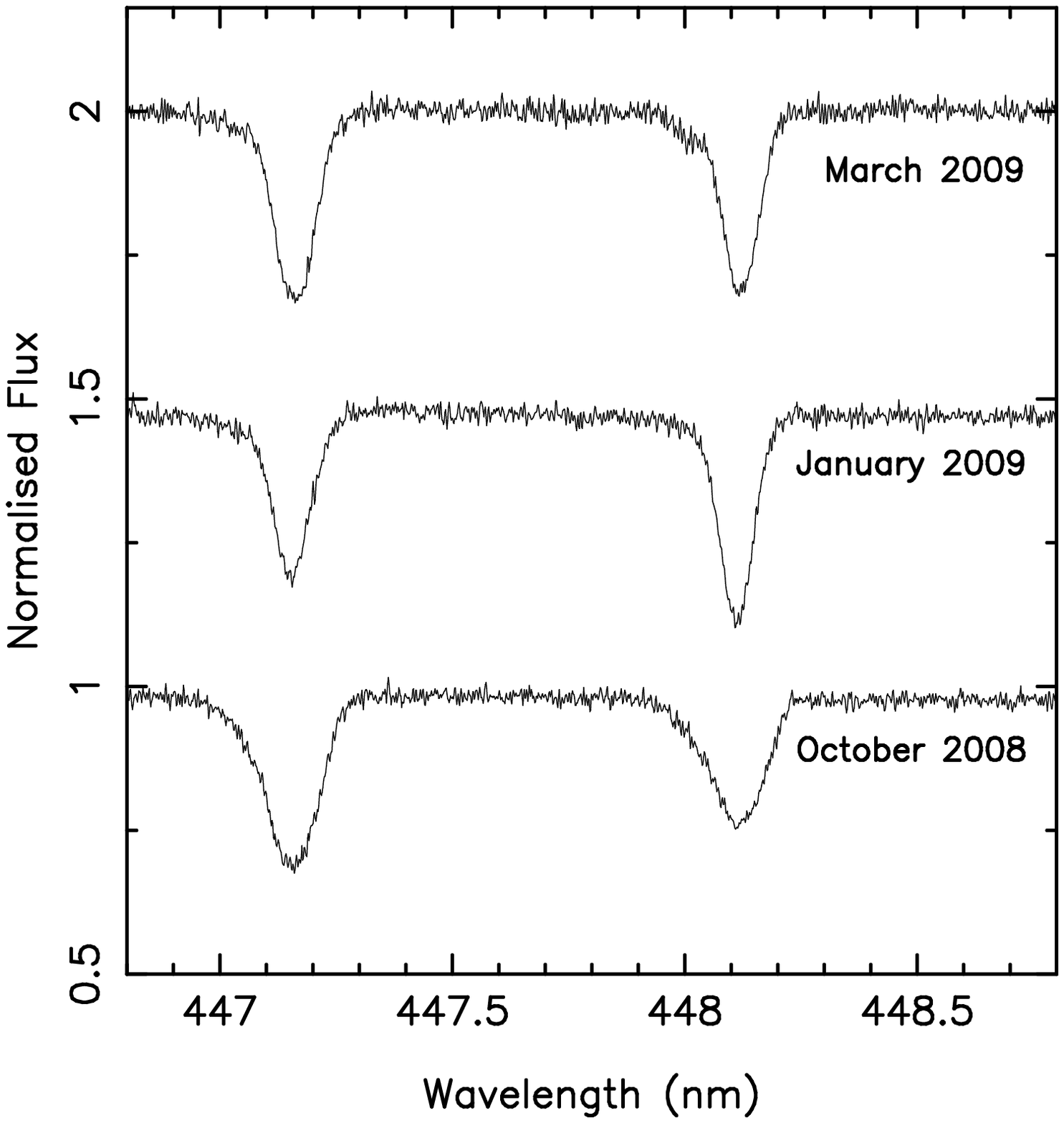}}}
\end{center}
\caption{Average profiles at three epochs
of H, He, and an Mg line of HD\,50064 (from left to right:
H$\alpha$, H$\beta$, H$\gamma$, \ion{He}{I} at 6678\AA, and \ion{He}{I} at
4471\AA, as well as \ion{Mg}{II} at 4481\AA).}
\label{spectra}
\end{figure*}

The B-type 
supergiant HD\,50064 ($V$ mag of 8.21) has not been studied in detail.  
Its spectral
type assignments range from B1Ia (Jacoby \& Hunter 1984) to B6Ia (Blanco et al.\
1970).  It was monitored by CoRoT, whose performance not only delivers two
orders of magnitude better precision than any ground-based photometry, but, even
more importantly for supergiant stars, also guarantees uninterrupted data during
several months. This is essential for progress in understanding 
massive evolved stars, because 
ground-based data for supergiants have so far suffered
severely from very low duty cycles.

The oscillations of evolved massive stars known so far essentially come in two
flavours. Classical gravity mode oscillations with periods of a few days,
excited by the $\kappa\,$mechanism, have recently been found from space
photometry 
(Saio et al.\ 2006; Lefever et al.\ 2007a).  On
the other hand, theory predicts 
so-called strange modes with periods between roughly 10 and
100\,d in stars with masses above 40\,M$_\odot$.
These strange modes, which can be both radial and
non-radial in nature, are modes trapped in a cavity caused by a density
inversion in the very outer, highly non-adiabatic envelope of stars with a high
L/M ratio whose radiation pressure dominates over the gas pressure
(Glatzel \& Kiriakidis 1993; Saio et al.\ 1998; Glatzel et al.\ 1999; Dorfi \&
Gautschy 2000).  This type of oscillation has been claimed to be responsible
for the mass-loss episodes of luminous stars, such as luminous blue variables
(LBVs), e.g.\ Glatzel \&
Kiriakidis (1993), 
but observational proof of the occurrence of strange modes has
not been established so far.  Our data of 
HD\,50064 { suggests that the star undergoes 
a strange mode oscillation.}


\section{Data description}

\subsection{The CoRoT data }

HD\,50064 was observed by CoRoT during a long run in the anticentre direction
(LRa01) for 136.9 days.  It is the only hot supergiant among the
seismology targets so far.  The CoRoT light curve contains 319\,913
datapoints, with a time sampling of 32\,s, after deleting the
measurements suffering from hot pixels during the passage through the South
Atlantic Anomaly. 
In order to compare the space photometry behaviour of HD\,50064
with the one reported in the literature (Halbedel 1990),
we transferred the CoRoT fluxes  to { white-light} magnitudes.  

All seismology targets of LRa01 are subject to a small
downward trend of instrumental
origin.
{ In the case of HD\,50064,  the trend is clearly stronger and probably intrinsic
to the star. It was corrected for by a linear approximation.}
This detrended light curve is shown as the thick line in the upper panel of
Fig.\,\ref{lc}.  Large variations occur, with peak-to-peak values of $\sim
0.2\,$mag and with a time scale of about one month. This is compatible with the
scarce data in Halbedel (1990).  The light curve also reveals a sudden rise in
amplitude near day 62.

\subsection{Follow-up spectroscopy}

Halbedel (1990) took one spectrum of HD\,50064 and found a P\,Cyg H$\alpha$
profile with a V/R ratio of 0.23 and a maximum red emission peak value of 4.2
continuum units. A few low-resolution low signal-to-noise UV spectra taken in
1979 with
the IUE satellite are also available, {but they do not allow 
  quantitative estimates of the stellar parameters
to be derived}. 
We assembled 14 high-resolution \'echelle
spectra of the star with the CORALIE spectrograph attached to the 1.2m Euler
telescope in La Silla, Chile, at three epochs 
{ 
(5, 5, 4 spread over 6, 8, 6  nights in Oct.\,08, 
Jan.\,09, Mar.\,09, respectively),}
after the CoRoT data revealed the star's variability.  The integration time was
30\,min, leading to an S/N level of about 50. The usable parts of the spectrum
have wavelengths between 4000\AA\ and 7000\AA.

The average profile of some selected lines for the three epochs are shown in
Fig.\,\ref{spectra}. The Balmer lines point towards mass loss that
{
is variable on the same time scale as the CoRoT photometry. The
He and Mg lines are also variable on this time scale. We 
cannot exclude additional spectroscopic variability with periods below a day.
}

\subsection{Interferometry}

Given that the star is surrounded by circumstellar matter, as the Balmer lines
reveal, we observed HD\,50064 with the near-IR interferometric instrument
VLTI/AMBER (Petrov et al.\ 2007) during Belgian GTO time in March
2009\footnote{program ID 083.D-0028}. The measurement was performed on the
closed triangle A0-K0-G1, providing baselines of 90, 80, and 125\,m, 
respectively.
Our analyses show that the target was {\it not\/} resolved. Based on the spread
on the data, we find that the target's half-light radius must be less than 1
milli-arcsec. For the distance estimate of $\sog\,2900$pc (Halbedel 1990), this
upper limit of 1 milli-arcsec translates to 650 R$_\odot$ for the disk or a
disk-to-star flux ratio below 7\% in the covered wavelength range (1.2 --
2.5~$\mu$m) when assuming a simple model of a circumstellar disk around an
unresolved star.


\section{Analyses of the data}

\subsection{Modelling of the CoRoT light curve}

\begin{figure}[ht!]
\begin{center}
\rotatebox{270}{\resizebox{4.5cm}{!}{\includegraphics{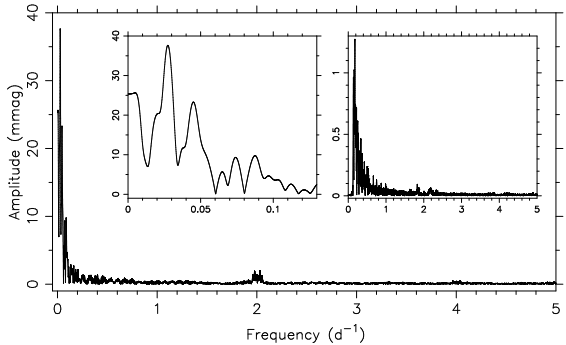}}}
\end{center}
\caption{Amplitude spectrum of the CoRoT data of HD\,50064, up to 5\,d$^{-1}$
(57.87$\mu$Hz).  At higher frequencies, only harmonics of the satellite orbit of
6184\,s occur, with an amplitude below 1\,mmag.  The insets show an enlarged
section at low frequencies (left) and the amplitude spectrum of the residuals
after prewhitening with the dominant frequency and six of 
its (sub)harmonics (right).}
\label{scargle}
\end{figure}

A first look at the detrended light curve immediately revealed 
large non-sinusoidal variations in amplitudes of several hundredth of a
magnitude, with periodicity of tens of days (Fig.\,\ref{lc}).  The
data is thus highly oversampled. 
We
binned the light curve by averaging 101 consecutive data points.  All the
results we list are for this binned light curve, containing 3167 points and
having a Rayleigh limit of 0.0073\,d$^{-1}$. The reference epoch we adopted is
$t_0={\rm HJD}\ 2454391.970406$.

The amplitude spectrum of the binned data is shown in Fig.\,\ref{scargle}. All
intrinsic variability due to the star occurs below 1\,d$^{-1}$. The highest peak
occurs at 0.027\,d$^{-1}$ (0.313$\mu$Hz), corresponding to a period of 37\,d,
and reaches an overall amplitude of 38\,mmag. We checked explicitly that the
same result for the frequency is obtained for the full (un)detrended unbinned
light curve.  A harmonic fit with this frequency, with a constant value for the
amplitude and phase of each harmonic, is not appropriate, even when considering
numerous harmonics (dashed-dotted line in Fig.\,\ref{lc}).  This is due to the
change of variability near day 62 (see Fig.\,\ref{lc}), which cannot be an
instrumental effect, as it does not occur for the other asteroseismology targets
of LRa01.  The variance in the light curve increases with a factor 1.6 after day
62.

We cut the binned data set into two parts, one until day 61.7 (set A) and
another one after it (set B).  The dominant frequencies for set A are 0.015 and
0.030\,d$^{-1}$, while we find 0.024\,d$^{-1}$ for set B. These frequencies
differ less or only slightly more than the Rayleigh limit from each other.
Given the larger variability in set B, this data dominates the
periodogram of the overall light curve, which leads us to conclude that there is
only one clear periodicity in the CoRoT data, corresponding with a frequency of
0.027\,d$^{-1}$, but with an amplitude and/or phase change near day 62. A
separate harmonic fit with 0.027\,d$^{-1}$ fixed for sets A and B and including
the (sub)harmonics $nf_i$, for $n=1/2,1,3/2,2,5/2,3$ is shown as a 
dashed line in
Fig.\,\ref{lc}. It leads to an amplitude and phase switch from 28\,mmag to
46\,mmag and from $\phi=0.623$ to $\phi=0.560$ (2$\pi$ radians).
Such a fit is not worse than one for which we allow the frequency to be
optimised for sets A and B (dotted line in Fig.\,\ref{lc}).
The residuals in the bottom panel
of Fig.\,\ref{lc} show variability at the 1\,mmag level 
(inset in Fig.\,\ref{scargle}) caused by the higher harmonics of
0.027\,d$^{-1}$, while no additional independent frequencies were found.

We conclude that the CoRoT light curve of HD\,50064 can be characterised by a
single period of 37\,d with a sudden amplitude increase of a factor 1.6,
which occurs once on a time scale of 137\,d.

\subsection{Spectroscopic behaviour}
\begin{figure}[ht!]
\begin{center}
\rotatebox{270}{\resizebox{5.5cm}{!}{\includegraphics{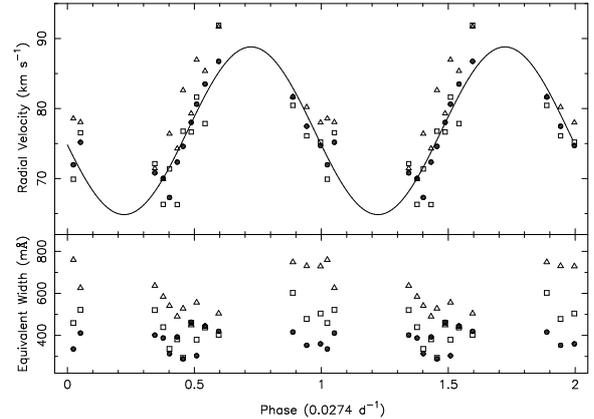}}}
\end{center}
\caption{Phase diagrams of the radial velocity and equivalent width of the three
spectral lines \ion{He}{I} 4471\AA\ (squares), \ion{He}{I} 6678\AA\
(triangles), and \ion{Mg}{II} 4481\AA\ (filled dots). Each measurement
occurs twice for visibility purposes. }
\label{vrad}
\end{figure}

Our spectroscopic data of HD\,50064 is limited to 14 spectra of moderate S/N,
but it spans 169\,d (Rayleigh limit of 0.006\,d$^{-1}$).  We computed the
centroid velocities (see Aerts et al.\ 1992) for the least blended spectral
lines, as well as their Fourier transform.  This led to {\it clear confirmation
of the photometric period}.  Phase diagrams for the \ion{He}{I} 4471\AA,
\ion{He}{I} 6678\AA, and \ion{Mg}{II} 4481\AA\ lines are shown in
Fig.\,\ref{vrad}.  We find a consistent radial-velocity (RV) amplitude of
12$\pm$2\,km\,s$^{-1}$ from these three lines. This amplitude is similar to the
one of B-type radial pulsators along the main sequence (e.g., Aerts \& De Cat
2003).
Large equivalent width (EW) variations occur for the three considered lines,
with relative peak-to-peak values near 50\% (see bottom panel of
Fig.\,\ref{vrad}). These also reveal the 37\,d period for the two He lines. The
phase difference between the EW and RV variations amounts to 
$277^\circ \pm 14^\circ$ for
\ion{He}{I} 4471\AA\ and $246^\circ \pm 10^\circ$ 
for \ion{He}{I} 6678\AA. Adiabatic
pulsation modes would give rise to values of $90^\circ$ or $270^\circ$,
depending on the spectral line. Deviations thereof are usually interpreted as
non-adiabatic effects (De Ridder et al.\ 2002).

We computed an average spectrum for the three epochs after shifting each
spectrum to the centroid velocity of the \ion{Mg}{II} 4481\AA\ line
(Fig.\,\ref{spectra}).  This reveals a range in centroid velocities of up to
50\,km\,s$^{-1}$ for lines formed at different optical depths. The V/R ratio of
H$\alpha$ is $\simeq 0.44$ in Oct.\,08, $\simeq 0.56$ in Jan.\,09, and $\simeq
0.73$ in Mar.\,09.  Both 
H$\alpha$ and H$\beta$ evolve from a double-peaked emission
profile to a P\,Cygni profile in that period, while H$\gamma$ transforms from an
inverse P\,Cygni profile in Oct.\,08 to a modest P\,Cygni profile in Jan.\,09
and an absorption profile in Mar.\,09, and  
H$\delta$ is always in absorption.  We
interpret this behaviour in terms of radial motion of the atmosphere in the line
forming region of H$\alpha$ and H$\beta$ in Oct.\,08 while at the same time
material recedes towards the stellar centre deeper in the atmosphere where
H$\gamma$ and the He and metal lines are formed.  This situation evolves to an
expanding upper atmosphere in Jan.\ and Mar.\,09, which are separated by two
pulsation cycles.

We determined \teff\ and \logg\ from the EW of \ion{Si}{ii} and \ion{Si}{iii}
lines, as well as \ion{He}{i} lines, by comparing with the predictions for the
extended grid of FASTWIND model atmospheres {\tt BSTAR06} in Lefever et al.\
(2007b).  The absence of \ion{He}{ii} and \ion{Si}{iv} lines places a clear
upper limit to \teff.  Restricting the models
to the solar Si abundance (Asplund et al.\
2009) led to $\teff\simeq 13500$\,K, $\logg\simeq 1.5$, where the differences
for the three epochs were less than the typical uncertainties of $1000\,$K and
0.2 dex. The EW of the Balmer lines led to $\log Q = \log [\dot{M} (v_\infty
R)^{-1.5}] \simeq -11.5$.  On the other hand, the peak heights of H$\alpha$
range from 1.5 in this study to 4 in Halbedel (1990) and this leads to the rough
estimate $\dot{M}\sim 10^{-5}$M$_\odot$\,yr$^{-1}$ as an average mass-loss rate.
This, combined with the value of $v_\infty \simeq 100\,$km\,s$^{-1}$ derived
from the blue wings of the H$\alpha$ profiles, gives a radius estimate of
$\simeq 200\,$R$_\odot$ and a luminosity $\log(L/L_\odot)\simeq 6.1$.

\section{Interpretation}

{ The 37\,d period found in the CoRoT photometry and in the spectroscopy is
  compatible with the radial fundamental mode for the parameters of HD\,50064,
  assuming $M\sim 45\,$M$_\odot$ (e.g., Lovy et al.\ 1984). The behaviour of the
  Balmer lines is hard to explain in terms of non-radial gravity modes as found
  in other B supergiants (Lefever et al.\ 2007a). While the overall morphology
  of the light curve with the sudden amplitude changes resembles the
  theoretically predicted modes by Dorfi \& Gautschy (2000), their predicted
  periods are an order of magnitude too short, and they did not give rise to 
mass-loss episodes. We thus suggest that HD\,50064 is subject to a
  radial strange mode oscillation.  }

The spectrum of HD\,50064 shows a strong 
resemblance to that of LBVs with moderate
mass loss, e.g.\ HD\,160529 (Stahl et al.\ 2003). We thus suggest that
HD\,50064 is on its way to that stage, by building up a circumstellar envelope
while pulsating. The main conclusion of our work is that its
pulsation mode is clearly connected with its variable mass loss.
With only three epochs of spectroscopy, we cannot make a detailed comparison of
the spectroscopic and pulsational behaviour, but we suggest a coordinated action
for long-term photometric and high S/N spectroscopic follow-up observations in
order to understand the detailed behaviour of the (strange) mode(s) and the
relation with the mass loss of this supergiant.

\begin{acknowledgements}
CA and KL acknowledge discussions with Joachim Puls on stellar winds.
The research leading to these results received funding from the
ERC under the European Community's 7th Framework
Programme (FP7/2007--2013)/ERC grant agreement n$^\circ$227224 (PROSPERITY), as
well as from the Research Council of K.U.Leuven and
from the Belgian Federal Science Policy Office.
\end{acknowledgements}

\end{document}